Microstructure and Fe-vacancy ordering in the KFe$_x$Se$_2$ superconducting system

Z. Wang[1], Y. J. Song[1], H. L. Shi[1], Z.W. Wang[1], Z.Chen[1], H. F. Tian[1], G. F. Chen[2], J.G. Guo[1], H. X. Yang[1] and J. Q. Li[1]*

[1] Beijing National Laboratory for Condensed Matter Physics, Institute of Physics, Chinese Academy of Sciences, Beijing 100190, P. R. China

[2] Department of Physics, Renmin University of China, Beijing 100872, P. R. China

Structural investigations by means of transmission electron microscopy (TEM) on K$_{0.8}$Fe$_x$Se$_2$ and KFe$_x$Se$_2$ with 1.5 ≤ x ≤ 1.8 have revealed a rich variety of microstructure phenomena. Materials with 1.5 ≤ x ≤ 1.6 often show a superstructure modulation along the [310] zone-axis direction, and this modulation can be well interpreted by the Fe-vacancy order which likely yields a superstructure phase of K$_2$Fe$_4$Se$_5$. The superconducting KFe$_x$Se$_2$ (1.7 ≤ x ≤ 1.8) materials contain clear phase separation in particular along the **c**-axis direction recognizable as visible parallel lamellae in the crystals, this fact suggests that the superconducting phase could have the Fe-vacancy disordered state.

Since the discovery of high-temperature superconductivity in the quaternary ZrCuSiAs-type (so called the 1111-phase) Fe-based oxypnictide LaFeAsO (F doped) with critical temperature ($T_c$=26 K) in 2008 [1], Superconductivity in the PbO-type α-FeSe$_x$ and related materials has been extensively investigated [2]. More recently, the superconductivity with $T_c$~33 K has been found in FeSe layered compound $K_{0.8}Fe_{1.7}Se_2$ and $Cs_{0.8}(FeSe_{0.98})_2$ respectively [3-5], which inspires wide interests on study of this kind of superconducting materials. Experimental studies revealed no structural phase transition occurring over the temperature ranging from 60 to 300 K [6]. The effect of varying Fe-content on transport properties of $KFe_xSe_2$ [7] and substitution on K site, such as $(Tl,K)Fe_xSe_2$ and $RbFe_2Se_2$ [8-10], have been also investigated. Measurements of optical properties of $K_{0.8}Fe_xSe_2$ demonstrate that the superconductivity is in close proximity to an antiferromagnetic semiconducting phase [11]. In $KFe_xSe_2$, the potassium sheets are intercalated between the FeSe layers stacking along **c**-axis direction, as discussed for $TlFe_xS_2$ and $TlFe_xSe_2$ [12, 13], and they all have an antiferromagnetic structure at low temperatures. Experimental investigations of Mössbauer spectroscopy and neutron diffraction demonstrated the presence of Fe vacancies in this system, and the Fe-deficiency and related ordered states are considered playing a critical role for understanding of the microstructure and physical properties. In present work, we performed a transmission-electron-microscopy (TEM) study on $K_{0.8}Fe_xSe_2$ and $KFe_xSe_2$ with $1.5 \leq x \leq 1.8$. The microstructure features and superstructures corresponding to different Fe-vacancy orders are reported.

Both single crystal and polycrystalline samples were used in present study. $Fe_{1+y}Se$ was firstly synthesized as precursor by reacting Fe powder with Se powder at 750°C for 20 hours. K pieces and $Fe_{1+y}Se$ powder were put into an alumina crucible with nominal compositions as $KFe_xSe_2$ ($1.5 \leq x \leq 2.0$). The single crystals used in present study were synthesized by method as reported in previous publications [3]. Specimens for TEM observations were prepared by peeling off a very thin sheet of a thickness around several tens microns from the single crystal and then milling by Ar ion. Microstructure investigations were performed on a FEI Tecnai-F20 TEM equipped with double-tilt cooling holder.

The structural and physical properties of all $K_{0.8}Fe_xSe_2$ and $KFe_xSe_2$ samples used for our TEM study were well characterized. The x-ray diffraction measurements demonstrate that these samples in general have a tetragonal basic structure with lattice parameters *a*=*b*=3.913 Å, *c*=14.10 Å, and the space group of *I*4/*mmm* (No. 139). It is similar with the crystal structure of the 122 phase in the known Fe-based superconducting systems, such as $KFe_2As_2$, $BaFe_2As_2$ and $SrFe_2As_2$ [14-16]. In order to understand the microstructure features in correlation with the Fe-deficiency in present system, we performed an extensive structural investigation by means of selected-area electron diffraction and high-resolution TEM observations.

We firstly focus our study on the $K_{0.8}Fe_xSe_2$ ($1.5 \leq x \leq 1.6$) which shows a semiconductor-like resisitivity from room temperature down to 4K. Figures 1(a)-(d) show a series of electron diffraction patterns from different areas taken along the relevant [001], [100], [1-30] and [001] zone-axis directions. The main diffraction spots with relatively strong intensity can be well indexed by the known tetragonal structure in consistence with the x-ray diffraction results [3]. On the other hand, the most striking structural phenomenon revealed in our TEM observations is the appearance of a series of superlattice spots following with the main diffraction spots, as clearly illustrated in Figs. 1(a), (c) and (d). Careful examination reveals that these satellite spots in general are clearly visible in the *a\*-b\** plane of reciprocal space and can be characterized by a unique modulation wave vector, $\mathbf{q_1}$ = (3/5, 1/5, 0) as also confirmed by the observations along the [1-30] zone-axis direction as illustrated in Fig.1(c). A careful analysis on chemical composition for this superstructure phase suggests the presence of a Fe-deficient $K_2Fe_4Se_5$ phase (i.e. $K_{0.8}Fe_{1.6}Se_2$) with the lattice parameters of $a_s = b_s$ = 6.15 Å and $c_s$ = 14.13 Å. The Fe valence state in this superstructure phase is $Fe^{2+}$. Occasionally, other kinds of superstructures can be also observed in certain areas as discussed in the following context.

Sometimes, two sets of superstructure reflections appear around each basic Bragg spot as shown in Fig. 1(d) in which two sets of superstructure reflections are indicated by $\mathbf{q_1}$ and $\mathbf{q_2}$, respectively. These two sets of superstructure spots are considered to originate from the domains where the superstructure vectors are twinning related respect to one another. This kind of twin domain was previously observed and reported in a variety of superstructure systems [17, 18], and the $\mathbf{q_1}$ and $\mathbf{q_2}$ could respectively appear in different areas or different layers in a crystal.

Fig. 2(a) shows the high-resolution TEM image taken from a $K_{0.8}Fe_{1.6}Se_2$ single-crystalline sample, indicating a well-defined 122-type layered structure [19].

This image taken along the [100] zone-axis direction was obtained from a relative thicker region of a crystal. The K atom positions are therefore recognizable as bright dots. The Se and Fe atoms with a distance of 1.56 Å is not resolvable in this image, but yield bright contrast between K layers.

We now go on to discuss the superstructure as illustrated in Fig. 1 in correlation with the Fe-deficiency in present superconducting system. According to previous neutron diffraction study for $TlFe_xSe_2$ ($1.5<x<2$) [12] and careful analysis of our TEM data, we herein interpret this superstructure phase by a Fe-vacancy order along the [310] zone-axis direction. In order to directly observe the ordered arrangement of the Fe vacancies in the crystals, we carried out the high-resolution TEM investigation on the superstructure along several relevant directions. Fig. 2(b) shows a high-resolution TEM image taken from thin crystal, in which the ordered behavior as visible periodic features within the *a-b* plane can be clearly read out. Image calculations based on the proposed superstructure model as shown in Fig. 2(b) were performed by varying the crystal thickness from 5 to 10 nm and the defocus value from -40 to -100 nm. A calculated image with a defocus value of -57nm and a thickness of 9 nm is superimposed onto the image, and it appears to be in good agreement with the experimental image.

It is also noted that the superstructure with lattice parameter *2×2* $d_{110}$ as discussed for $KFe_{1.5}Se_2$ in previous literature [20] appears frequently in certain areas. Fig. 3(a) shows a high resolution image revealing this kind of superstructure as observed in a crystal. This TEM image is taken in an area with a thickness estimated to be 8nm, clearly displaying a $KFe_{1.5}Se_2$ crystal with visible superstructure features. Careful examinations suggest that the ordered state of the Fe vacancies adopt a well defined *2×2* supercell corresponding with the structural model of Fig. 3(b). Our study on this superlattice suggests the appearance of another phase of $K_2Fe_3Se_4$ (i.e. $KFe_{1.5}Se_2$). Further investigation on the structural and physical properties of this phase is in progress.

It is remarkable to note that the increase of the Fe-concentration could result in the appearance of superconductivity in this system, e.g the $KFe_{1.8}Se_2$ shows clear superconducting transition at *Tc* ~33K. Actually, experimental studies show that the Fe-concentration affects visibly the electrical transport and structural properties of this layered system. For instance, a clear insulator-metal transition often appears in samples with $1.6 \leq x \leq 1.8$ and disappears for high Fe-concentration [7]. This fact suggests that the Fe-vacancy and the ordered states could play a critical role for understanding the physical properties in $KFe_xSe_2$ .

For better understanding of the microstructure properties in the superconducting phase, we have focused our attention on the superconducting samples with nominal composition of $KFe_xSe_2$ ($1.7 \leq x \leq 1.8$) which in general show sharp superconducting transitions at around 33 K. TEM observations indicated that all samples with the nominal compositions of $KFe_{1.8}Se_2$ have apparently inhomogeneous microstructure and clear phase separation, in particular along the **c**-axis direction.

Fig.4 shows the high resolution TEM images obtained from a superconducting sample. It is clearly recognizable that the superconducting crystals in general have the same tetragonal basic structure and a clear superstructure within the **a-b** plane as discussed in the above context. However, clear changes of microstructure and superstructure properties can be commonly observed. Fig. 4(a) shows a high-resolution TEM image revealing the presence of complex microstructures. Careful analysis on the structural properties in relatively large domain size (>100nm) suggests a similar superstructure as discussed in the $KFe_{1.8}Se_2$ sample with $q_1$ = (3/5, 1/5, 0), but the structural inhomogeneity and changes of ordered behaviors occur visibly along $q_1$ direction as shown in the Fig. 4(b) in which a layer without the Fe-vacancy order is indicated by an arrow. Moreover, it is also noted that superstructure in superconducting crystals almost disappears in certain domains (the area B in Fig. 4(a)), as shown in Fig. 4(c). Though some contrast anomalies arising possibly from Fe vacancies can be seen in this area, the crystal actually has a basic tetragonal lattice without a long rang Fe-vacancy order.

Fig. 4(d) shows a high-resolution TEM image illustrating the phase separation along the **c** axis direction in the superconducting sample. It is clearly recognizable that the additional Fe ions could result in the appearance of Fe disordered layers in the superconducting crystals and yield the intergrowth of structural lamellae with the Fe-vacancy order state(OS) and disorder state (DOS) along the **c** axis direction. Our observations suggest that the superconducting phase in present system could have the Fe-vacancy disordered state as illustrated in Fig. 4(d). Moreover, in the superconducting sample $KFe_{1.8}Se_2$, we can also see a superstructure with periodicity of $q_3$ = (3/4, 1/4, 0) coexisting with $q_1$ = (3/5, 1/5, 0) in certain crystals. This fact suggests that the alteration of Fe-concentration in $KFe_xSe_2$ materials could yield not only superconductivity but also clear microstructure changes.

In conclusion, microstructure analysis of $K_{0.8}Fe_xSe_2$ and $KFe_xSe_2$ with $1.5 \leq x \leq 1.8$ reveals a rich variety of structural phenomena in correlation with the presence of Fe-deficiency in this layered system. A clear superstructure have been commonly observed with a modulation wave vector of $\mathbf{q_1}= (3/5, 1/5, 0)$. This superstructure can be well interpreted by the Fe-vacancy order within the *a-b* plane and suggests the presence of a $K_2Fe_4Se_5$ phase in this system. The increase of Fe-concentration in the $KFe_xSe_2$ materials could results in clear alternations of microstructure properties in superconducting materials. Structural phase separation and the complex superstructures commonly appear in the superconducting samples as observed by high-resolution TEM investigations, these facts suggest that the superconducting phase could have the Fe-vacancy disordered state.

**Acknowledgments**

 We would like to thank Prof X.L,Chen and Prof H.H,Wen for their fruitful discussion. This work is supported by the National Science Foundation of China, the Knowledge Innovation Project of the Chinese Academy of Sciences, and the 973 projects of the Ministry of Science and Technology of China.

Figure captions

Fig. 1. Electron diffraction patterns of $K_{0.8}Fe_{1.6}Se_2$ taken along (a) the [001], (b) the [100] and (c) the [1-30] zone-axis directions, respectively, showing the superstructure reflections within the $a*$-$b*$ reciprocal plane. (d) Electron diffraction pattern taken from a twinning area for superstructure.

Fig. 2. High-resolution TEM images of $K_{0.8}Fe_{1.6}Se_2$ taken along (a) the [100] and (b) the [001] zone-axis directions, respectively, displaying the atomic structural features of $K_{0.8}Fe_{1.6}Se_2$. In Fig. 2(a) the inset displays a simulated image of the sample with thickness of 40nm, a defcous of -12nm and a spherical aberration of 1.2mm. The inset of Fig. 2(b) shows a simulated image of the sample with thickness of 9nm, a spherical aberration of 1.2mm and a defocus of -57nm.

Fig. 3. (a) A high-resolution TEM image reveals the superstructure with supercell of *2×2* $d_{110}$ in a crystal of $KFe_{1.5}Se_2$. The ordered state of the Fe-vacancy adopts a well defined *2×2* supercell corresponding with the structural model of Fig. 3(b).

Fig. 4. (a-c) High-resolution TEM image shows inhomogeneous microstructure and complex domains in a crystal of $KFe_{1.8}Se_2$ [001]; (d) TEM observations along the [1-30] zone axis direction showing the phase separation along the **c**-axis direction.

Fig1

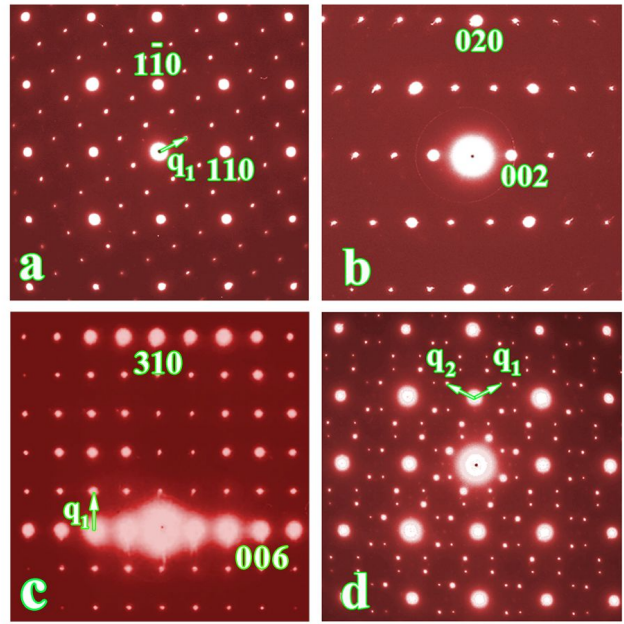

Fig 2

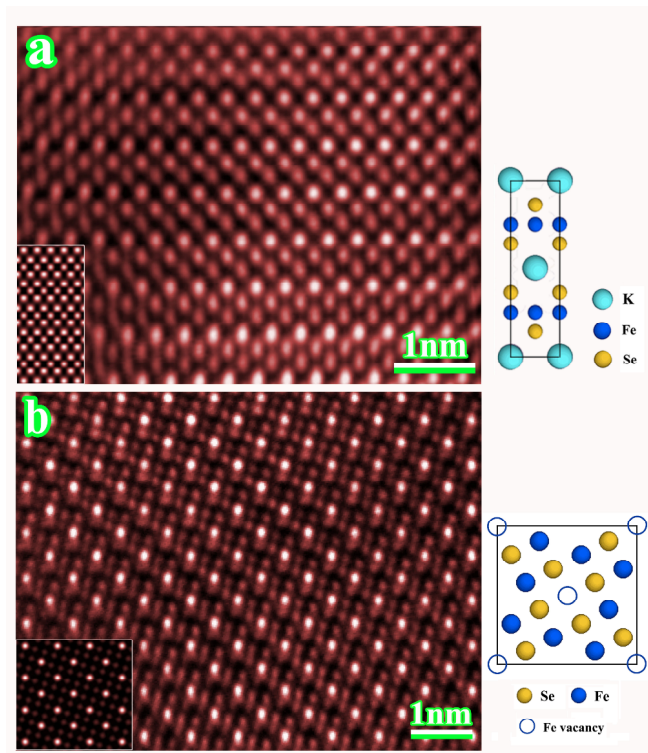

Fig.3

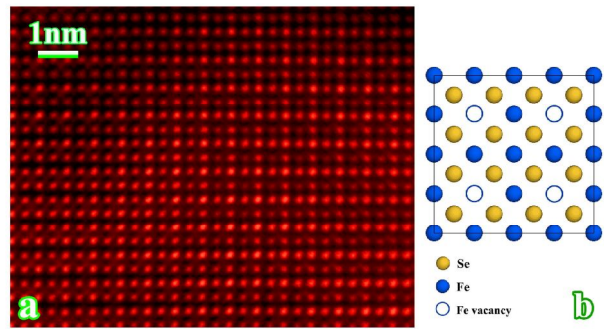

Fig4

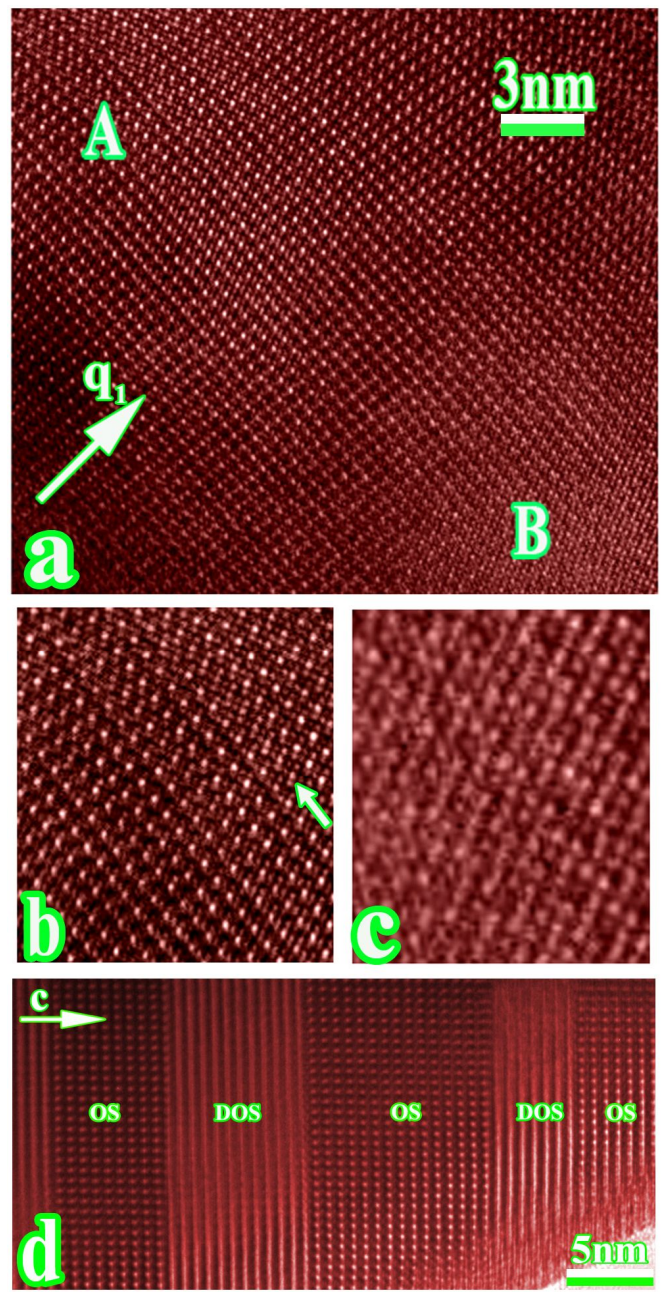